\documentclass[%
reprint,
superscriptaddress,
 amsfonts,
 amsmath,
 amssymb,
 aip,
 apl,
]{revtex4-1}

\usepackage{graphicx}

\usepackage{bm}
\usepackage{hyperref}

\usepackage{listings}
\usepackage[dvipsnames]{xcolor}
\usepackage{appendix}

\usepackage[utf8]{inputenc}
\usepackage[T1]{fontenc}
\usepackage{verbatim}

\DeclareRobustCommand{\uvec}[1]{{%
	\ifcsname uvec#1\endcsname
		\csname uvec#1\endcsname
	\else
		\bm{\hat{\mathbf{#1}}}%
	\fi
}}

\newcommand{\imi}{\mathrm{i}}
\newcommand{\eue}{\mathrm{e}}

\newcommand{\diff}{\mathop{}\!\mathrm{d}}

\begin{document}

\title{Thermal-noise Limits to the Frequency Stability of Burned Spectral Holes}

\author{M. T. Hartman}
\altaffiliation[Currently at: ]{{L'Institut Fresnel, 13013 Marseille, France}\\{Electronic mail: \href{mailto:michael.hartman@fresnel.fr}{michael.hartman@fresnel.fr}}}
\affiliation{LNE-SYRTE, Observatoire de Paris, Universit\' e PSL, CNRS, Sorbonne Universit\' e,\\ 75014 Paris, France}

\author{N. Wagner}
\affiliation{Institut für Halbleitertechnik, Technische Universität Braunschweig, Hans-Sommer-Str. 66,\\ 38106 Braunschweig, Germany}
\affiliation{Laboratory for Emerging Nanometrology, Langer Kamp 6a-b,\\ 38106 Braunschweig, Germany}

\author{S. Kroker}
\affiliation{Institut für Halbleitertechnik, Technische Universität Braunschweig, Hans-Sommer-Str. 66,\\ 38106 Braunschweig, Germany}
\affiliation{Laboratory for Emerging Nanometrology, Langer Kamp 6a-b,\\ 38106 Braunschweig, Germany}
\affiliation{Physikalisch-Technische Bundesanstalt, Bundesallee 100,\\ 38116 Braunschweig, Germany}

\author{S. Seidelin}
\affiliation{Univ. Grenoble Alpes, CNRS, Grenoble INP and Institut N\' eel,\\ 38000 Grenoble, France}

\author{B. Fang}
\email{bess.fang@obspm.fr}
\affiliation{LNE-SYRTE, Observatoire de Paris, Universit\' e PSL, CNRS, Sorbonne Universit\' e,\\ 75014 Paris, France}

\date{20 June 2025}

\begin{abstract}
Techniques in frequency stabilization of lasers to fixed-spacer optical cavities have advanced to the point where the ultimate frequency stabilities are limited by thermal noise for standard cavity configurations at room temperature.  In early experiments, laser stabilization via spectral-hole burning (SHB) has been shown to be a promising alternative.  In this letter, we explore the thermal-noise limits to frequency stability in spectral holes and compile known material parameters for a typical system used in SHB experiments ($\mathrm{Eu^{3+}}$doped $\mathrm{{Y_{2}}{Si}{O_{5}}}$) to make numerical estimates for the fundamental thermal-noise induced frequency instability in spectral-holes for the liquid-helium temperature and dilution temperature cases.  These efforts can guide the design of future SHB experiments and clarify which important material parameters remain to be measured.
\end{abstract}

\maketitle


\section*{\label{sec:Introduction}Introduction}

Spectral-hole burning (SHB) is a technique for producing persistent narrow transmissive spectral lines, or `spectral holes', within an inhomogeneously broadened absorption band\cite{moerner1988persistent}.  In practice, within a doped crystal, the broadened absorption band is provided by random spectral displacement of the narrow homogeneous absorption lines of the dopant ions or atoms caused by crystal matrix deformation at the doping-locations.  Selecting a system for which the ground state has multiple hyperfine levels allows for transmissive lines to be `burned' (or photo imprinted) into the absorption band by using a pump laser to deplete one of the hyperfine levels which is resonant with the pump laser frequency.  The result is a powerful technique which allows the production of an arbitrary spectral pattern at the desired working frequencies.  With this capability, SHB has been leveraged by researchers for diverse applications in fields such as quantum memory and information processing\cite{NILSSON2005393, Hedges2010, Bussieres2014, walther2015:PhysRevA.92.022319, Zhong2017, Maring2017, Laplane2017},  ultra-wideband spectral analysis\cite{Schlottau:05,Berger:16}, quantum opto-mechanics\cite{Molmer2016:PhysRevA.94.053804, seidelin2019:PhysRevA.100.013828, PhysRevLett.126.047404}, and laser frequency stabilization\cite{PhysRevB.62.1473, Julsgaard:07, PhysRevLett.107.223202, Thorpe2011,Gobron:17, PhysRevLett.114.253902, Galland:20}.

Regardless of their application, experiments involving SHB are concerned, to one degree or another, with the frequency noise in the spectroscopic measurement of the spectral hole.  This is directly evident in applications to laser frequency stabilization, for which SHB has been proposed as an alternative to optical cavities as a frequency reference. In this context, sources of frequency noise can be categorized into two groups: the first group, detection noise (shot noise, electronic noise, etc), limits the ability to resolve the spectral hole frequency. While the requirement to use low probe laser power in SHB accentuates this group, it is a technical noise source which has been mitigated by innovative detection schemes\cite{Galland:20,PhysRevA.104.063111, Lin:23}.  The second group consists of noise sources that shift the frequency of the spectral line itself.  A subset of this group is the intrinsic frequency instability due to thermal-noise sources; these constitute a fundamental physical limit and are the focus of this letter.

Currently, state of the art ultra-stable lasers reference their frequency to linear optical cavities, which consist of two partially reflective mirrors spaced a distance, $d$, apart.  An incident spatially mode-matched laser beam will produce a build-up of intracavity laser power when the resonance condition is met, that is, when the cavity length is an integer multiple, $N$, of the half wavelengths, $\frac{\lambda}{2}$, of incident light: $d = N\frac{\lambda}{2}$.  Typically a feedback scheme\cite{Drever1983, Black2001:10.1119/1.1286663, PhysRevD.92.022004} uses the intracavity field to produce an error signal to control the incident laser frequency, $\nu_0$, locking the laser to one of the resonant modes which are spaced at intervals of the free spectral range, $\mathrm{FSR}=\frac{c}{2d}$, where $c$ is the speed of light in vacuum.  The result is a reference where the fractional frequency fluctuations are proportional to cavity length changes: $\frac{\delta{\nu}}{{\nu_0}} = -\frac{\delta{d}}{d}$. 

By contrast, frequency fluctuations in spectral holes can result from phenomena which produce energy shifts in the transition.  The significance of externally driven effects on the frequency stability in SHB has motivated the characterization of spectral-hole line shifts due to several types of environmental perturbations. These characterizations have focused on a particularly promising medium, $\mathrm{Eu^{3+}}$ doped into a $\mathrm{{Y_{2}}{Si}{O_{5}}}$ crystal matrix (Eu:YSO), whose long lifetime in the hyperfine states\cite{oswald_characteristics_2018} and long optical coherence time\cite{equall_ultraslow_1994} at cryogenic temperatures present desirable spectroscopic properties.  One such important effect is the line shift resulting from a change in the electric field interacting with the dipole moment of the ion transition.  This can result from either an applied external field (Stark effect)\cite{Zhang2020:doi:10.1063/5.0025356} or from external mechanical stress causing a deformation of the crystal lattice around the ion\cite{Thorpe2011, Zhang2020aPhysRevResearch.2.013306, Galland2020:PhysRevApplied.13.044022}.  A second significant effect is a change in phonon density which results in an energy shift in the transition as predicted in a two-phonon scattering process\cite{PhysRevB.68.085109}.  The result is manifest as temperature dependent line shift\cite{PhysRevB.68.085109, Thorpe2011, PhysRevLett.107.223202, Thorpe_2013,Zhang2023:PhysRevA.107.013518,Lin2024}.  Making use of these previous studies' characterizations of spectral-hole response to external forces, as well as fresh measurements of material properties, this letter estimates the fundamental frequency-stability of spectral-holes due to intrinsic thermal-noise.

Thermal noise presents a fundamental physical limit in numerous precision measurement experiments.  Broadly speaking, thermal noise describes the fluctuation of a measurable coordinate as a result of internal thermal-energy driven motion.  In the context of the development of cavity-based ultra-stable lasers, thermal noise is manifest as a fluctuation in the apparent surface position of cavity mirrors, as calculated by application of the fluctuation-dissipation theorem\cite{PhysRevD.57.659,BRAGINSKY19991,PhysRevD.65.102001,Harry_2002}.  This is already well known as the limiting frequency-noise source, typically inducing a fractional-frequency instability of order low $10^{-16}$ up to  $10^{-15}$ in standard optical-cavity-referenced lasers\cite{young_visible_1999,ludlow_compact_2007}, depending on cavity geometry, where the models \cite{Numata2004:PhysRevLett.93.250602} are well verified by measurements\cite{notcutt_contribution_2006}.  This limitation has driven the exploration of SHB as an alternative frequency reference for the development of ultra-stable lasers\cite{PhysRevB.62.1473, Julsgaard:07, PhysRevLett.107.223202, Thorpe2011, PhysRevLett.114.253902, Galland:20}, yet, we have not found systematic analysis of the thermal noise processes in SHB in the literature.  As the mechanisms producing frequency shifts in spectral holes are fundamentally different from those in optical cavities, we here calculate thermal noise in SHB.  In this work, we derive expressions for thermal-noise induced frequency-fluctuations via two mechanisms, thermomechanical and thermodynamical noise, for a typical sample geometry used in SHB: a rectangular parallelepiped crystal doped with rare-earth ions.  We then apply known material and spectroscopic parameters to provide representative numerical estimates of thermal-noise driven fractional frequency instability in SHB.

\section*{\label{sec:Thermomechanical_Noise}Thermomechanical Noise}
Thermomechanical noise, or Brownian thermal noise, describes a displacement noise that is the result of mechanical energy dissipation, driven by random thermal motion, through the internal mechanical loss angle, $\phi$, due to internal friction.  We propose a coupling mechanism of thermomechanical noise to spectral-hole frequency-noise through stress induced deformation of the crystal lattice.  An important material parameter in this coupling, the frequency sensitivity to mechanical-stress, $\left({K}_D=\frac{\delta{\nu}}{\delta{P}}\, \left[\frac{\mathrm{Hz}}{\mathrm{Pa}}\right]\right)$, has been previously characterized in Eu:YSO along two ($D=1,2$) of the three dielectric axes using externally applied uni-axial pressures\cite{Galland2020:PhysRevApplied.13.044022}. Using this, we can write a vector of spectral-hole fractional-frequency sensitivity to mechanical stress,
\begin{equation}
    {{\overrightarrow{k}}} = \begin{bmatrix} {{k}_{1}}\\ {{k}_{2}}\\{{k}_{3}} \end{bmatrix} = \frac{1}{\nu_0}\begin{bmatrix} {K_{1}}\\ {K_{2}}\\{K_{3}} \end{bmatrix}.\label{eqn:SHBStressSensitivity}
\end{equation}

To aid in our calculation of thermal noise we make use of the fluctuation dissipation theorem (FDT). Developed by Callen et al.\cite{Callen1952:PhysRev.86.702}, the FDT describes systems in which a perturbing force $F$, which displaces coordinate $x$, is irreversibly dissipated into thermal energy.  In general, the FDT gives the power spectral density (PSD) of fluctuations in $x$, $S_x$, or, conversely, the PSD of fluctuations in $F$, $S_F$, over a spectrum of angular frequencies $\omega$,
\begin{align}
    S_x\!(\omega) &= \frac{4k_\mathrm{B}T_0}{\omega}\Im{\left[\chi(\omega)\right]},\label{eqn:FDT1}\\
    S_F\!(\omega) &= \frac{4k_\mathrm{B}T_0}{\omega}\Im{\left[\frac{1}{\chi(\omega)}\right]},\label{eqn:FDT2}
\end{align}
as a function of the internal thermal energy (as given by the Boltzmann constant, $k_\mathrm{B}$, and temperature, $T_0$) and the imaginary part of the complex mechanical susceptibility of $x$ to $F$,
\begin{equation}
    \chi\!(\omega) = \frac{x(\omega)}{F(\omega)},
\end{equation}
which contains resistive and dissipative components. 

\begin{figure}[t]
    \includegraphics{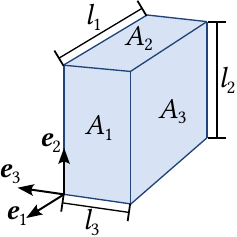}
    \caption{\label{fig:crystal_diagram}Dimensional diagram of a crystal used in SHB.}
    \vspace{-1.5ex}
\end{figure}

We examine the case of a rectangular parallelpiped crystal matrix of volume $V_\mathrm{x}=l_{1}l_{2}l_{3}$, illustrated in FIG. \ref{fig:crystal_diagram}, where the lengths, $l_j$, along axes $\bm{e}_j$, and orthogonal crystal cross-sectional areas, $A_j$, are given by the corresponding dimensional matrices,
\begin{equation}
    \mathbf{M}_{l}=\begin{bmatrix} l_{1} & 0 & 0 \\ 0 & l_{2} & 0 \\ 0 & 0 & l_{3} \end{bmatrix},\quad
    \mathbf{M}_\mathrm{A}=\begin{bmatrix} l_{2}l_{3} & 0 & 0 \\ 0 & l_{1}l_{3} & 0 \\ 0 & 0 & l_{1}l_{2} \end{bmatrix}. 
\end{equation}
Given a material with isotropic complex Young's modulus $\widetilde{E} = E(1-\imi\phi)$ and Poisson's ratio $\sigma$, we can write the mechanical compliance matrix,
\begin{equation}
    \mathbf{M}_\mathrm{c}=\frac{1}{\widetilde{E}}\begin{bmatrix} 1 & -\sigma & -\sigma \\ -\sigma & 1 & -\sigma \\ -\sigma & -\sigma & 1 \end{bmatrix}, \label{eqn:young}
\end{equation}
which allows us to express the complex mechanical susceptibility of the crystal,
\begin{equation}
    \chi=\mathbf{M}_{l}\mathbf{M}_\mathrm{c}\mathbf{M}^{-1}_\mathrm{A}.
\end{equation}
We can incorporate \eqref{eqn:SHBStressSensitivity} into the FDT \eqref{eqn:FDT2} to write the PSD of fractional-frequency noise contribution along each dielectric axis due to {\em thermomechanical noise},
\begin{align}\label{eqn:PSD_ff_thermomechanical}
    \overrightarrow{S}_{\!\frac{\delta{\nu}}{\nu_0}}\!(\omega)&=\frac{4k_\mathrm{B}T_0}{\omega} \Im{\left(\chi^{-1}\right)} \left(\mathbf{M}^{-1}_\mathrm{A}\right)^2 \left({{\overrightarrow{k}}}\odot{{\overrightarrow{k}}}\right)\\
    &=\frac{4k_\mathrm{B}T_0}{\omega}\frac{E\phi}{(1-2\sigma)(1+\sigma)}\frac{1}{l_1l_2l_3} \nonumber\\
    &\qquad\times\begin{bmatrix} {\frac{1}{l_2l_3}\left({k}^2_{1}l_2l_3(1-\sigma) +  {k}^2_{2}l_1l_3\sigma  +  {k}^2_{3}l_1l_2\sigma\right)}\\ 
    {\frac{1}{l_1l_3}\left({k}^2_{1}l_2l_3\sigma +  {k}^2_{2}l_1l_3(1-\sigma) +  {k}^2_{3}l_1l_2\sigma\right)}\\ 
    {\frac{1}{l_1l_2}\left({k}^2_{1}l_2l_3\sigma +  {k}^2_{2}l_1l_3\sigma +  {k}^2_{3}l_1l_2(1-\sigma)\right)} \end{bmatrix}. \nonumber
\end{align}
where $\odot$ denotes a Hadamard product, an element-wise multiplication. 

\section*{\label{sec:Thermo-spectral_Noise}Thermo-spectral Noise}

A second category of thermal noise, a thermodynamic noise, describes the measurable effects from intrinsic temporal variations of temperature over a probed volume known from statistical thermodynamics\cite{landau2013statistical}. Such a class of fundamental noise has been identified and characterized in laser cavity experiments, where temperature fluctuations give rise to cavity-length fluctuations via the thermoelastic and thermorefractive effects in the cavity mirror substrates\cite{BRAGINSKY19991,Cerdonio2001:PhysRevD.63.082003,Evans2008:PhysRevD.78.102003} and coatings\cite{BRAGINSKY2000303,BRAGINSKY2003244,Evans2008:PhysRevD.78.102003}.  By contrast, spectral-hole temperature dependence is the result of phonon-scattering driven spectral line shifts with a general $T^4$ trend\cite{PhysRevB.68.085109}. This mechanism, aside from anomalous localized deviations\cite{Lin2024}, produces a general $T^3$ trend in frequency sensitivity to temperature $\left(\Gamma = \left. \frac{\diff{\nu}}{\diff{T}} \right\rvert_{T_0} \, \left[\frac{\mathrm{Hz}}{\mathrm{K}}\right]\right)$ driving experiments to lower temperatures for performance gains.  In the case of SHB, we are interested in the temperature variations in the volume of crystal being probed by a laser beam of radius $r_\mathrm{b}$. 

Historically, for most cases ($\approx 10\,\mathrm{cm}$ optical cavities at room temperature), the time scales of interest for temperature fluctuations are much shorter than the system's thermal time constant, $\tau_\mathrm{th}=\frac{r_\mathrm{b}^2\rho c_\mathrm{p}}{\kappa}$, set by the radius of the cylindrical probed area, $r_\mathrm{b}$, and the density, $\rho$, specific heat capacity at constant pressure, $c_\mathrm{p}$, and thermal conductivity, $\kappa$, of the probed material; in those cases, an adiabatic approximation, which assumes that temperature variations occur within confines of the probed region, is valid\cite{BRAGINSKY19991}.  However, in the case of small beam sizes or high thermal diffusivity($\alpha = \frac{\kappa}{\rho c_\mathrm{p}}$), temperature fluctuations diffuse over a distance larger than the probe beam during the measurement times of interest, requiring a more general expression for the PSD of temperature fluctuations, $S_T$, within a volume.  This has been solved for the case of a beam reflecting from a surface\cite{Cerdonio2001:PhysRevD.63.082003}, and, in a case more applicable to SHB, in transmission through an optic, where we follow Braginsky \& Vyatchanin\cite{BRAGINSKY2004} to a solution:
\begin{equation}\label{eqn:PSD_T_thermodynamical}
    S_T\!(\omega) = \frac{k_\mathrm{B}T_0^2}{\pi l_3\kappa}\int_0^\infty\frac{u\eue^{-u}}{\left(\omega\tau_\mathrm{th}\right)^2 + u^2} \,\diff{u}.
\end{equation}
From \eqref{eqn:PSD_T_thermodynamical}, we can write the PSD of fractional-frequency noise due to temperature fluctuation driven spectral-hole line shifts, which we refer to as {\em thermo-spectral noise},
\begin{equation}\label{eqn:PSD_ff_thermodynamical}
    S_{\!\frac{\delta{\nu}}{\nu_0}}\!(\omega) = \gamma^{2}S_T\!(\omega),
\end{equation}
in terms of the fractional-frequency temperature sensitivity, $\gamma=\frac{1}{\nu_0}\Gamma$.

\section*{\label{sec:NumericalEstimates}Numerical Estimates}

\begin{table*}[ht]
\begin{center}
\begin{tabular}{ p{3cm} p{3cm} p{3cm} p{3cm} p{2cm} p{2cm} }
    \hline			
    \hline
     
    & Value
    & Value
    & Value
    & 
    & 
    \\
    Parameter 
    & (Room Temp)
    & ($4\,\mathrm{K}$)
    & ($300\,\mathrm{mK}$)
    & Units
    & Reference
    \\
    \hline
    $r_\mathrm{b}$
    & $3.5$
    & -
    & -
    & $\mathrm{mm}$ 
    & -
    \\
    ${l_1}\times{l_2}\times{l_3}$
    & $8\times8\times4$
    & ${\dagger}$
    & ${\dagger}$
    & $\mathrm{mm^3}$
    & \cite{Galland2020:PhysRevApplied.13.044022} 
    \\
    $\rho$
    & $4400$
    & ${\dagger}$
    & ${\dagger}$
    & $\mathrm{kg\,m^{-3}}$ 
    & \cite{Mirzai2021}
    \\
    $E$
    & $150$
    & $158$
    & $158$
    & $\mathrm{GPa}$ 
    & \cite{Mirzai2021, Wagner:24}
    \\    
    $\sigma$
    & $0.26$
    & ${\dagger}$
    & ${\dagger}$
    & -
    & \cite{Mirzai2021}
    \\
    $c_\mathrm{p}$
    & 
    & $2\!\times\!{10}^{-2}$
    & $1\!\times\!{10}^{-5}\:\triangle$ 
    & $\mathrm{J\,{kg}^{-1}\,{K}^{-1}}$
    & \cite{CpManuscriptInPrep} 
    \\
    $\kappa$
    & 
    & $40$
    & $0.02\:\square$
    & $\mathrm{W\,{m}^{-1}\,{K}^{-1}}$
    & see Appendix
    \\    
    $\nu_0\:{\ast}$
    & $516.85$
    & -
    & -
    & $\mathrm{THz}$
    & \cite{Galland2020:PhysRevApplied.13.044022}
    \\
    $K_{1}\:{\ast}$
    & N/A
    & $46$  
    & ${\ddagger}$
    & $\mathrm{Hz}\,\mathrm{Pa}^{-1}$
    & \cite{Galland2020:PhysRevApplied.13.044022}
    \\
    $K_{2}\:{\ast}$
    & N/A
    & $-19$  
    & ${\ddagger}$
    & $\mathrm{Hz}\,\mathrm{Pa}^{-1}$
    & \cite{Galland2020:PhysRevApplied.13.044022}
    \\
    $K_{3}\:{\ast}$
    & N/A
    & $\:{\Diamond}$
    & ${\ddagger}\:{\Diamond}$
    & $\mathrm{Hz}\,\mathrm{Pa}^{-1}$
    & \cite{Galland2020:PhysRevApplied.13.044022}
    \\
    $\Gamma\:{\ast}$
    & N/A
    & $23\,000$
    & $5$ 
    & $\mathrm{Hz\,K^{-1}}$
    & \cite{Lin2024}
    \\
    \hline
    \hline
\end{tabular}
\caption{\label{tab:shb_material_parameters} Eu:YSO material and spectroscopic parameter values used in the estimation of thermal noise in SHB {\footnotesize \\
$\ast$ Value used for spectroscopic site 1, the site typically used in laser stabilization experiments.
\\
$\dagger$ Value at room temperature was used for estimates, it is not expected to vary greatly going to cryogenic temperatures.\cite{ekin_experimental_2014}
\\
$\ddagger$ Value unknown at $300\,\mathrm{mK}$.  For calculations at $300\,\mathrm{mK}$ we use $4\,\mathrm{K}$ values.
\\
$\Diamond$ Value unknown for dielectric axis 3, we set this value to the larger of other two axis values ($46\,\mathrm{Hz}\,\mathrm{Pa}^{-1}$) for the numerical estimates.
\\
$\triangle$ Values at 4\,K are measured, whereas values at 300\,mK are model projected from measurements performed in the range from 380 mK to 2.4 K.\cite{CpManuscriptInPrep}\\
$\square$ Values at 4\,K are measured, whereas values at 300\,mK are model projected. See Supplemental Material.
}}
\vspace{-3ex}
\end{center}
\end{table*}

In order to provide an order-of-magnitude estimate of the fundamental stability limit in current SHB experiments, here we use the expressions we previously derive to compute thermal-noise induced frequency instability using the typical material and experimental parameters of current and past SHB laser stabilization experiments.  Namely, we follow the experimental setup as presented by the SHB group at LNE-SYRTE\cite{Gobron:17, Galland:20}, which utilizes a rectangular cuboid YSO crystal\cite{Galland2020:PhysRevApplied.13.044022} of dimension ${8}\times{8}\times{4}\,\mathrm{mm}^3$ doped with $\mathrm{Eu}^{3+}$ at a concentration of 0.1 atomic \%. However other SHB laser stabilization experiments use similarly dimensioned samples (for example a 0.005\% Er:YSO cuboid of ${5}\times{6}\times{1}\,\mathrm{mm}^3$ at Montana State University\cite{PhysRevB.63.155111}; a 1\% Eu:YSO cylinder of length $5\,\mathrm{mm}$ and $10\,\mathrm{mm}$ in diameter at NIST\cite{Thorpe2011}; a 0.5\% Eu:YSO cuboid of ${5}\times{5.5}\times{10}\,\mathrm{mm}^3$ at University of Düsseldorf\cite{oswald_characteristics_2018}).  While characterizations of material and spectroscopic properties of Eu:YSO at cryogenic temperatures are incomplete, we have compiled some known property values at various temperatures in TABLE \ref{tab:shb_material_parameters}, which are used in all our numerical estimates. We also note that doping concentration would generally affect many material properties.  Variations in mechanical properties due to doping concentration are expected to be relatively small, with a less that 2\% variation in the Young's modulus predicted throughout a large range of doping concentrations (0\% to 25\%)\cite{Mirzai2021}.  However, it should be noted the variations of thermal properties, especially at low temperatures, can be much larger\cite{slack_thermal_1971}, and while this study uses values for thermal properties specifically measured for 0.1\% Eu:YSO (the concentration used in configuration we follow, SHB experiment at LNE-SYRTE), care is advised to the reader before extending the use of these values to other doping concentrations in other experiments.  Furthermore, a narrow range of Eu$^{3+}$ concentration has been shown to be practical for modern ultra-stable laser applications (0.1at.\% in\cite{Galland:20}, 0.5at.\% in\cite{oswald_characteristics_2018}, and 1at.\% in\cite{Thorpe2011}), likely hitting a sweet spot between sufficient Eu$^{3+}$ for a reasonable signal-to-noise ratio and interaction between neighbouring Eu$^{3+}$ ions that may reduce coherence, limiting the need for further consideration of doping concentration effects on material properties in the application of SHB to ultra-stable lasers.

\begin{figure*}[t]
    \includegraphics{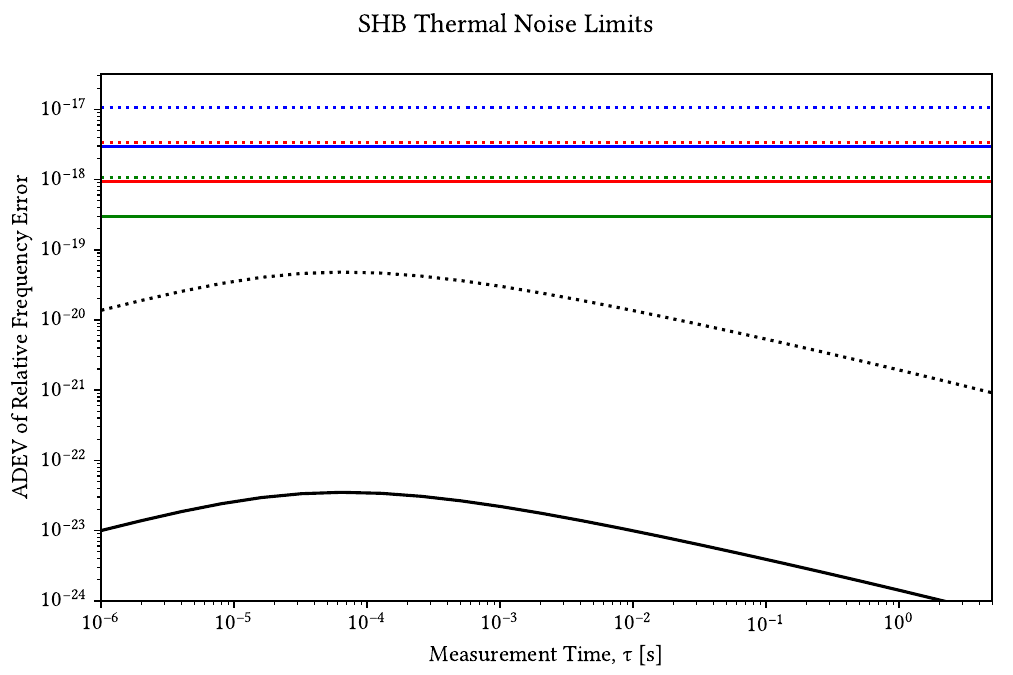}
    \caption{\label{fig:ADEV_ThermalNoise} The calculated Allan deviation (ADEV) due to thermal noise versus measurement times at temperatures $T=300\,\mathrm{mK}$ (solid lines) and $T=4\,\mathrm{K}$ (dotted lines). The colored lines correspond to Brownian noise for several possible loss angles of $\phi=10^{-4}$ (blue), $\phi=10^{-5}$ (red) and  $\phi=10^{-6}$ (green), whereas the black lines describe the thermo-spectral noise limit. For the thermo-spectral noise over short time-scales ($\tau<\tau_\mathrm{th}$), the slope proportional to $\sqrt{\tau}$ in ADEV ($\frac{1}{f^2}$ in PSD) corresponds to the adiabatic region, where temperature fluctuations are confined within volumes less than the total probed volume.  At longer time-scales, thermal diffusion of the temperature fluctuations reduces the thermo-spectral effect.
    }
\end{figure*}

For thermomechanical noise, we take the interaction between the orthogonal axes to be coupled via Poisson's ratio, and, in the case described in TABLE \ref{tab:shb_material_parameters} (i.e., the LNE-SYRTE SHB experiment, where $l_{1}=l_{2}=L$ and $l_{3}=l$), \eqref{eqn:PSD_ff_thermomechanical} becomes  $S_{\!\frac{\delta{\nu}}{\nu_0}}\!(\omega)  =  \overrightarrow{S}_{\!\frac{\delta{\nu}}{\nu_0}} (\omega) \cdot \overrightarrow{e}_j$:
\begin{align}
    S_{\!\frac{\delta{\nu}}{\nu_0}}\!(\omega) &= \frac{4k_\mathrm{B}T_0}{\omega}\frac{E\phi}{(1-2\sigma)(1+\sigma)}\frac{1}{V_\mathrm{x}}\Biggl(({k}^2_{1}+{k}^2_{2})\left(1+\sigma\frac{l}{L}\right)\nonumber\\
    &\qquad +{k}^2_{3}\left(1+\sigma\frac{2L-l}{l}\right)\Biggl).
\end{align}
The mechanical loss angle is known to be a function of temperature\cite{UCHIYAMA2000310}, though different materials show significantly different, to even inverse, proportionalities with temperature\cite{Vacher2005}.  Measurements of $\phi$ in YSO\cite{Wagner:24} have indicated decreasing internal mechanical losses with cooling, while not reaching the temperatures concerned here.  As such, in FIG. \ref{fig:ADEV_ThermalNoise}, we plot the Allan deviation (ADEV),
\begin{equation}\label{eqn:PSD2ADEV}
    \sigma_{\!\frac{\delta{\nu}}{\nu_0}}\!(\tau)=\sqrt{\frac{1}{\pi}\int_{0}^{\infty}{S_{\!\frac{\delta{\nu}}{\nu_0}}\!(\omega)}   \frac{\sin^4\!\left(\frac{1}{2}\tau\omega\right)}{\left(\frac{1}{2}\tau\omega\right)^2}\,\diff{\omega}}\:,
\end{equation}
of relative frequency instability due to thermomechanical noise over three orders of magnitude of $\phi$ (Green, Red, and Blue); other physical parameter values used for this plot can be found in TABLE \ref{tab:shb_material_parameters}.

For thermo-spectral noise, we use the spectral-hole temperature sensitivity $\gamma$ and the PSD of intrinsic temperature fluctuations $S_T\!(\omega)$ by numerical integration of \eqref{eqn:PSD_T_thermodynamical} to find the PSD of fractional-frequency fluctuations $S_{\!\frac{\delta{\nu}}{\nu_0}}\!(\omega)$ according to \eqref{eqn:PSD_ff_thermodynamical}. The Allan deviation is then evaluated by integrating $S_{\!\frac{\delta{\nu}}{\nu_0}}\!(\omega)$  with an appropriate kernel according to \eqref{eqn:PSD2ADEV}. We plot the resulting Allan deviation of thermo-spectral noise induced frequency instability in Black in FIG. \ref{fig:ADEV_ThermalNoise}.  

Additionally, while previous SHB experiments operated near liquid helium temperatures\cite{Thorpe2011, Galland:20} ($\approx 4\,\mathrm{K}$), a change to operating at dilution refrigerator temperatures ($<1\,\mathrm{K}$) has been strongly motivated\cite{Lin2024}.  As such, we plot each curve at temperatures $T=4\,\mathrm{K}$ (Dashed) and $T=0.3\,\mathrm{K}$ (Solid).

\section*{\label{sec:Discussion_and_Conclusions}Discussion and Conclusions}

In this letter, we have presented key figures of merit in the computation of the thermal-noise limits to spectral-hole frequency stability.  We find the thermal noise in Eu:YSO is dominated by thermomechanically-driven Brownian noise.  This term dominates over thermodynamically driven thermo-spectral noise, which is suppressed by a combination of the high thermal diffusivity of Eu:YSO at cryogenic temperatures as well as the low temperature sensitivity ($\Gamma$) of spectral-holes at dilution temperatures, particularly near a zero crossing discovered around $300\,\mathrm{mK}$ at spectroscopic site 1 in Eu:YSO\cite{Lin2024}.

We also note in passing that excitation induced line broadening and optically induced motion, such as the recently discovered piezo-orbital back action\cite{louchet-chauvet_strain-mediated_2023,louchet-chauvet_piezo-orbital_2023}, have not been taken into account in the current study due to the specific time scales relevant to experiments of SHB-based laser frequency stabilization in Eu:YSO. As demonstrated in our recent investigation\cite{lin_homogeneous_2025}, we obtain spectral holes with maximal dispersion when using low intensity pump beam (Gaussian beam of peak intensity $\sim 710$\,nW.cm$^{-2}$ and waist 3\,mm) and over several seconds of pumping time. With an optical coherence time of the order of ms\cite{equall_ultraslow_1994}, any excited ions would have already decayed into the ground state manifold by the time the frequency locking procedure starts. The probe beam, resonant with the spectral hole, interacts with the ions only dispersively and does not cause further direct excitations. As such, any effect related to rare earth ions in the optical excited states are not immediately relevant. 

While the fundamental stability of spectral holes is important in various experiments, we highlight the direct importance to laser frequency stabilization.  Ultra-stable lasers find an application in time and frequency metrology, where they are used in the probing of atomic transitions in optical lattice clocks.  In this context, the goal of reaching the fundamental performance limit of these optical clocks, the quantum projection noise, sets a target fractional-frequency instability of order $10^{-18}$ for 0.1 to 10\,s averaging time\cite{abdelhafiz2019guidelinesdevelopingopticalclocks}. For this purpose, when assuming a mechanical loss angle of order below $10^{-4}$, SHB in Eu:YSO is a compatible frequency reference for this application.  This assumption is realistic in the view of the first mechanical quality factor measurements in YSO\cite{Wagner:24}.

Here, we point out assumptions made and notable limitations to the estimates provided, namely missing material parameters.  In the area of thermo-spectral noise, we have measurements of $c_\mathrm{p}$ and $\kappa$ down to $2\,\mathrm{K}$.  Below $2\,\mathrm{K}$ we rely on a model\cite{gamsjager_low_2018} fit to the data, however this provides a reasonable value and is, anyway, the significantly smaller contribution to the total thermal noise. On the subject of thermomechanical noise, the sensitivity of spectral holes to stress, $K$, along dielectric axis 3 ($K_{3}$) has not been yet been measured, and would require modifications of our crystal and setup to be able to do so\cite{Galland2020:PhysRevApplied.13.044022}. Instead, in this letter, we substituted this value with the higher measured sensitivity of the other two axes to provide a conservative estimate, but the possibility of an exceptionally high sensitivity along this axis cannot be excluded. Furthermore, the other dielectric axes, $K_1$ and $K_2$, have only been measured around $4\,\mathrm{K}$. However, the mechanism for this coupling is deformation of the crystal lattice around the dopant ions due to external stress.  Since this mechanical deformation resulting from physical stress is governed by the rigidity of the crystal, which does not vary significantly at these temperatures (e.g. an indirect measurement of Young's modulus through the frequency of a mechanical resonance shows a saturation behaviour under $\sim 100$\,K in ref\cite{Wagner:24}), we do not expect $K$ to change significantly from $4\,\mathrm{K}$ to $300\,\mathrm{mK}$. Additionally, we have used an isotropic Young's modulus and Poisson's ratio for the YSO, a crystal with known anisotropy. However, this simplification remains valid for our order-of-magnitude estimation, as the Young's modulus for the individual axes are predicted to have a deviation of $< 15\%$ from the isotropic Young's modulus\cite{luo_theoretical_2014}.

We note that the experimental verification of such noise level would be possible if all other sources of noise are reduced to a level well below. This is, however, not the case at the present stage of development\cite{barbarat_guidelines_2025}, since technical noise such as that induced by residual vibrations of the cryogenic environment, often in the range of $\geq 10^{-16}$ projected fractional frequency instability at 1 s, must be properly handled.  This said, the closest measurement to date set an upper-bound for thermo-mechanical noise at $7.3\times{10}^{-17}$ in a similarly dimensioned SHB experiment at NIST\cite{PhysRevLett.111.237402}, remaining compatible with our thermal-noise calculation.

Nevertheless, the framework presented here provides key elements for the design of future SHB experiments using  Eu:YSO, as well as needs for further YSO characterizations.  Furthermore, despite incomplete knowledge of material properties, we believe the numerical estimates presented here reflect an accurate assessment of thermal-noise limits to fractational-frequency stability in present-day SHB experiments.

\section*{Appendix: Measurement of the thermal conductivity of 0.1\% Eu:YSO at cryogenic temperatures}

\begin{figure}[htb!]
    \includegraphics[width=0.83\linewidth]{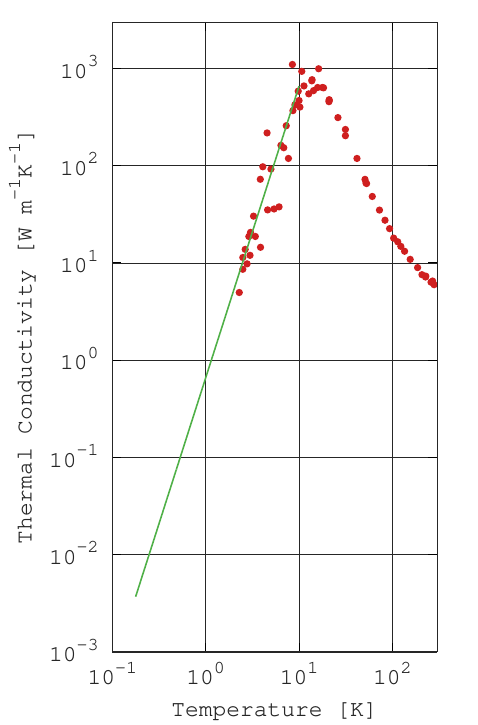}
    \caption{\label{fig:kappa_log_plot}Thermal conductivity of $0.1\%$ Eu:YSO.}
\end{figure}

The thermal conductivity of $0.1\%$ Eu:YSO above $2\,\mathrm{K}$ is measured using the Physical Property Measurement System of Quantum Design at the Plateforme Mesures Physiques à Basses Températures, Sorbonne Université. Their data is plotted in FIG. \ref{fig:kappa_log_plot} (red points).  Values below $2\,\mathrm{K}$ were estimated assuming an ideal dielectric crystal $\kappa=\beta{T}^3$ fit at $T< 10\,\mathrm{K}$ resulting in $\beta=0.65\pm0.08\,\mathrm{Wm^{-1}K^{-4}}$. 

\section*{Acknowledgements}

We thank Alban Ferrier for preparing the Eu:YSO samples and David Hrabovsky for carrying out the measurements of the thermal conductivity. We acknowledge funding from Ville de Paris Emergence Program, the Région Ile de France DIM C’nano and SIRTEQ, the LABEX Cluster of Excellence FIRST-TF (ANR-10-LABX-48-01) within the Program ``Investissement d'Avenir'' operated by the French National Research Agency (ANR), the Deutsche Forschungsgemeinschaft (DFG, German Research Foundation) under Germany's Excellence Strategy—EXC-2123 QuantumFrontiers—390837967, the 20FUN08 NEXTLASERS project from the EMPIR program co-financed by the Participating States and from the European Union’s Horizon 2020 research and innovation program, and the UltraStabLaserViaSHB (GAP-101068547) via Marie Skłodowska-Curie Actions (HORIZON-TMA-MSCA-PF-EF) in the European Commission Horizon Europe Framework Programme (HORIZON). We also acknowledge ANR funding via the grant ANR-24-CE47-1190. 


\textbf{Data availability.} The data that support the findings of this study are available from the corresponding author upon reasonable request.

\bibliography{references}

\end{document}